\documentclass[10pt]{article}
\pdfoutput=1

\usepackage{amsmath,amsfonts,amssymb}
\usepackage{enumerate}
\usepackage{epsfig}
\usepackage{color}
\usepackage{textcomp}
\usepackage{wasysym}
\usepackage{cite}
\usepackage{url}

\usepackage{subfigure}
\usepackage{amsthm}

\usepackage{chngpage} 

\usepackage[T1]{fontenc}

\usepackage{etoolbox} 

\usepackage{stmaryrd} 

\newcommand{\layerindex}[1][]{ %
\ifstrequal{#1}{}{\ensuremath{\alpha} }{}
\ifstrequal{#1}{1}{\ensuremath{\alpha} }{}
\ifstrequal{#1}{2}{\ensuremath{\beta} }{}
\ifstrequal{#1}{3}{\ensuremath{\gamma} }{}
}
\newcommand{\layerindexvector}[1][]{ %
\ifstrequal{#1}{}{\ensuremath{\boldsymbol\alpha} }{}
\ifstrequal{#1}{1}{\ensuremath{\boldsymbol\alpha} }{}
\ifstrequal{#1}{2}{\ensuremath{\boldsymbol\beta} }{}
\ifstrequal{#1}{3}{\ensuremath{\boldsymbol\gamma} }{}
}

\newcommand{\nodeindex}[1][]{ %
\ifstrequal{#1}{}{\ensuremath{u} }{}
\ifstrequal{#1}{1}{\ensuremath{u} }{}
\ifstrequal{#1}{2}{\ensuremath{v} }{}
}

\newcommand{\aff}[1]{\textit{  #1 }}

\usepackage[font=footnotesize]{caption}

\newcommand{\comment}[2]{} 

\newcommand{\rev}[1]{{\color{black} #1}}

\newcommand{\muxviz}{\texttt{muxViz}}

\begin{document}

\title{MuxViz: A Tool for Multilayer Analysis and Visualization of Networks}


\author{
Manlio De Domenico\thanks{manlio.dedomenico@urv.cat}\\
\aff{Departament d'Enginyeria Inform\'atica i Matem\'atiques,}\\ \aff{Universitat Rovira I Virgili, 43007 Tarragona, Spain}
\and
Mason A. Porter\\
\aff{Oxford Centre for Industrial and Applied Mathematics,} \\ \aff{Mathematical Institute, University of Oxford, Oxford OX2 6GG, UK}\\ and \\ \aff{CABDyN Complexity Centre, University of Oxford, Oxford OX1 1HP, UK}
\and
Alexandre Arenas\\
\aff{Departament d'Enginyeria Inform\'atica i Matem\'atiques,}\\ \aff{Universitat Rovira I Virgili, 43007 Tarragona, Spain }
}

\maketitle



\begin{abstract}
{Multilayer relationships among entities and information about entities must be accompanied by the means to analyze, visualize, and obtain insights from such data. We present open-source software (\muxviz) that contains a collection of algorithms for the analysis of multilayer networks, which are an important way to represent a large variety of complex systems throughout science and engineering. We demonstrate the ability of \muxviz\, to analyze and interactively visualize multilayer data using empirical genetic, neuronal, and transportation networks. Our software is available at \url{https://github.com/manlius/muxViz}.}
{multilayer networks; software; visualization; multiplex networks; interconnected networks}
\\
2000 Math Subject Classification: 91D30, 05C82, 76M27
\end{abstract}

\section{Introduction}

Although the study of networks is old, the analysis of complex systems has benefited particularly during the last two decades from the use of networks to model large systems of interacting agents \cite{newman2010}. Such efforts have yielded numerous insights in many areas of science and technology\cite{kitano2002computational,de2002modeling,barabasi2004network,sharan2007network,beyer2007integrating,sporns2014contributions,colizza2006role,gomez2007dynamical,gomez2008spreading,eagle2009inferring,lazer2009life,balcan2009multiscale,kitsak2010identification,aral2012identifying,vespignani2012modelling}.

In the case of biological networks, connections among genes, proteins, neurons, and other biological entities can indicate that they are part of the same biological pathway or exhibit similar biological functions. Network representations focus on connectivity, and they have now become a paradigmatic way to investigate the organization and functionality of cells \cite{jeong2000large,jeong2001lethality,shen2002network,maslov2002specificity,tong2004global,guimera2005functional,rosenfeld2005gene,chen2006wiring,goh2007human,costanzo2010genetic}, synaptic connectivity \cite{van1996chaos,sporns2004motifs,buzsaki2004neuronal,sporns2004organization,mantini2007electrophysiological,bullmore2009complex,seeley2009neurodegenerative,bassett2011dynamic,nicosia2013remote,nicosia2013phase}, and more. There are also myriad applications to other types of systems (e.g., in sociology, transportation, physics, and more)\cite{Wasserman1994Social, boccaletti2006complex,newman2010,barthelemy2011spatial,Holme2012Temporal,kivela2013multilayer}.

In parallel, a large variety of computational techniques have been developed to analyze (and visualize) networks and the information that they encode. In biology, for example, such methods have become important tools for attempting to understand and represent cell functionality.
However, although the standard network paradigm has been very successful, it has a fundamental flaw: it forces the aggregation of multilayer information to construct network representations that include only a single type of connection between pairs of entities.  This can lead to misleading results, and it is becoming increasingly apparent that a more complicated representation is necessary \cite{kivela2013multilayer}.  

Recently, a novel mathematical framework to model and analyze multilayer relationships and their dynamics was developed \cite{mucha2010community,dedomenico2013mathematical}. In this framework, one represents the underlying network topology and interaction weights as a \emph{multilayer network}, in which entities can exhibit different relationships simultaneously and can exist on different ``layers''.  Multilayer networks can encode much richer information than what is possible using the individual layers separately (which is what is usually done).  This, in turn, provides a suitable framework for versatile and sophisticated analyses that have already been used successfully to reveal multilayer community structure \cite{mucha2010community} and to measure important nodes and the correlations between them \cite{dedomenico2013mathematical,dedomenico2013centrality,nicosia2013correlations,battiston2014structural}. However, to meet the requirements of an operational toolbox to be applied to the analysis of {complex} systems, it is of paramount importance to also develop open-source software to visualize multilayer networks and represent the results of analyzing such networks in a meaningful way.

Multilayer networks have already yielded fascinating insights and are experiencing burgeoning popularity.  For example, there have been numerous studies to attempt to understand how interdependencies (e.g., \cite{buldyrev2010catastrophic,brummitt2012suppressing}), other multilayer structures (e.g., \cite{lee2012correlated,radicchi2013abrupt,cardillo2013emergence,dedomenico2013centrality,cozzo2013clustering,nicosia2013correlations,cellai2013percolation,battiston2014structural}), dynamics (e.g., \cite{yaugan2012analysis,gomez2012evolution,cozzo2012stability,cozzo2013contact,dedomenico2014navigability}), and control (e.g., \cite{mario-review2010}) can improve understanding of complex interacting systems.  See the recent review article \cite{kivela2013multilayer} for extensive discussions and a thorough review of results.

\rev{The increasing use of more complicated network representations has yielded a new set of challenges: how should one visualize, analyze, and interpret multilayer data. 
Although there has been progress in numerous applications, many of the key results have concentrated on data from examples like social and transportation networks \cite{kivela2013multilayer}.
Multilayer analysis has rarely been exploited in the investigation of biological networks --- even though such a perspective is clearly relevant --- and we believe that the lack of appropriate software has contributed to this situation. For example,} in a recent study, the genetic and protein-protein interaction networks of \emph{Saccharomyces cerevisiae} were investigated simultaneously \cite{costanzo2010genetic} to uncover connection patterns. \rev{Costanzo \emph{et al} \cite{costanzo2010genetic} also reported that} genetic interactions have an overlap of 10--20\% with protein-protein interaction pairs, which is significantly higher than the $3\%$ overlap that they expected based on a random null model. This suggests that many positive and negative interactions occur between --- rather than within --- complexes and pathways \cite{costanzo2010genetic} and thereby gives an important example of how exploiting multilayer information might improve understanding of biological structure and functionality.

Although the aforementioned overlap is an indication of correlation between a pair of networks, the analysis of multilayer biological data would benefit greatly from techniques and diagnostics that are able to exploit multiplexity (i.e., multiple different ways to interact) in available information. 
\rev{As has been the case in several studies of social and technological systems \cite{mucha2010community,dedomenico2013centrality,nicosia2013correlations,battiston2014structural,dedomenico2014navigability}, the analysis of multilayer biological data would benefit greatly from techniques and diagnostics that are able to exploit, e.g.,  multiplexity (i.e., multiple different ways to interact) in available information.}


\section{Methods}

The primary contributions of the present work are to address the computational challenge of analysis and visualization of multilayer information by providing a practical methodology, and accompanying software that we call \muxviz, for the analysis and the visualization of multilayer networks. In Appendix\,\ref{supp:note:5}, we give technical details about the \muxviz\, software.


\subsection{Visualization}\label{supp:note:1}

\begin{figure}[!t]
\centering\includegraphics[width=16cm]{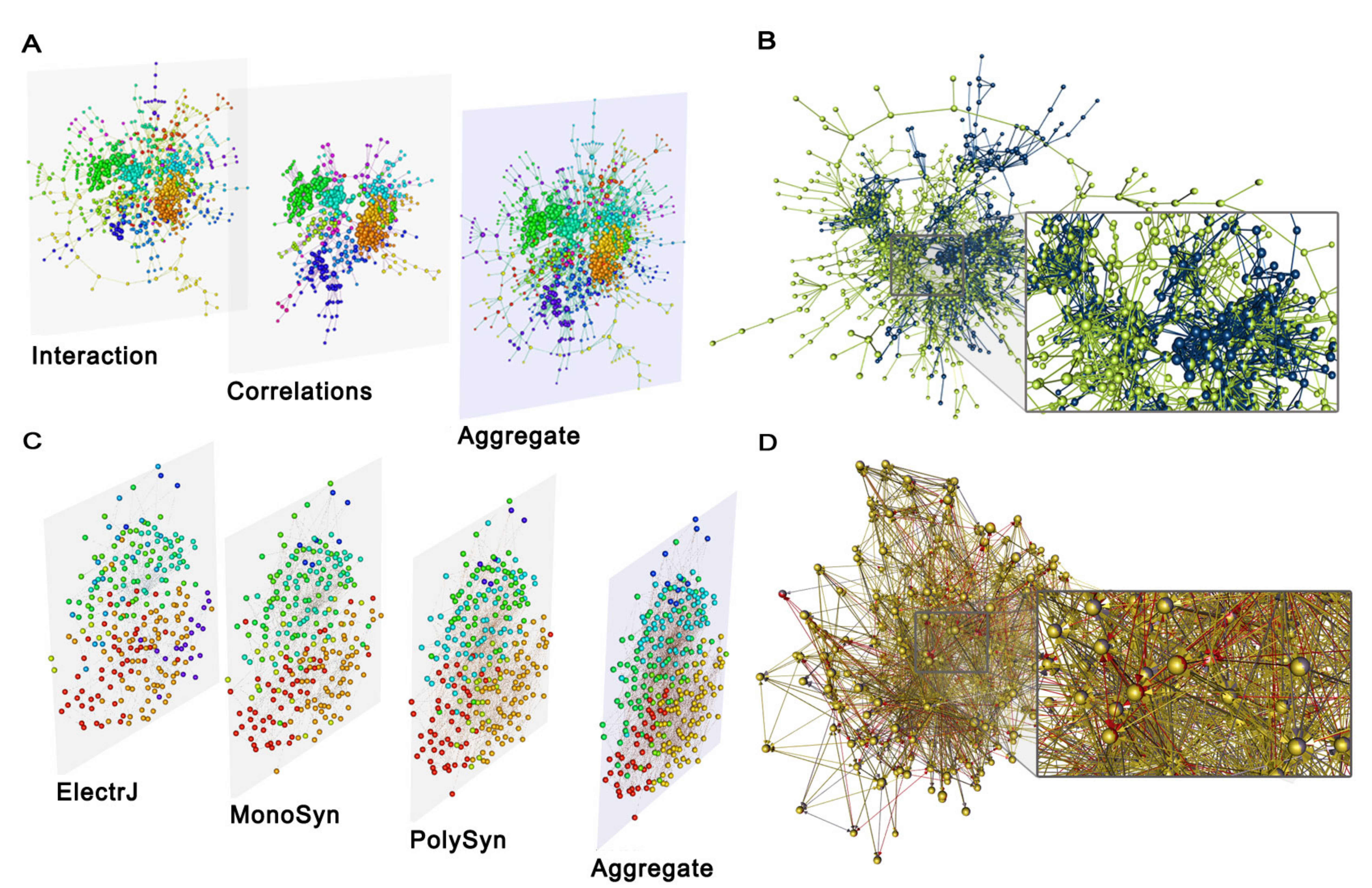} 
\caption{\textbf{Figure\,\ref{fig:fig1suppl}: Multilayer representations of genetic and neuronal networks}. \textbf{(A)} Multilayer representation, in which the layers correspond to interaction network of genes in \emph{Saccharomyces cerevisiae} (which was obtained via a synthetic genetic-array methodology) and a correlation-based network in which genes with similar interaction profiles are connected to each other. [The data comes from Ref.\,\cite{costanzo2010genetic}.] In the third layer, we show the corresponding aggregated network. In this visualization, the color of the nodes is their module assignment from multilayer community detection (see the text for further details). \textbf{(B)}. Representation of the same network as an edge-colored multigraph. \textbf{(C)} Multilayer and \textbf{(D)} edge-colored-multigraph representations of the \emph{Caenorhabditis elegans} connectome, where layers correspond to different synaptic junctions: electric (``ElectrJ''), chemical monadic (``MonoSyn''), and polyadic (``PolySyn''). [The data comes from Ref.\,\cite{chen2006wiring}.] In panels \textbf{B} and \textbf{D}, we color the nodes according to the layer to which they belong. If a node is part of multiple layers simultaneously, then we use an equal distribution of the corresponding colors for the node.}
\label{fig:fig1suppl}
\end{figure}

\rev{
In multilayer networks, nodes can exist in several layers simultaneously and entities that exist in multiple layers (such nodes have ``replicas'' on other layers) are connected to each other via interlayer edges. One can visualize a multilayer network in \muxviz\, either using explicit layers or as an edge-colored multigraph \cite{kivela2013multilayer}, in which edges are ``colored'' according to the different types of relationships between them (see Fig.\,\ref{fig:fig1suppl} for examples of genetic and neuronal multilayer networks).
}

\rev{
The \muxviz\, software focuses predominantly on ``multiplex networks'', which refer to networks with multiple  relational types and which are arguably the most important (and prevalent) type of multilayer network.  A large variety of systems in the biological, social, information, physical, and engineering sciences can be described as multiplex networks.  
In \muxviz, we consider two different types of interlayer connectivity: \emph{ordinal} and \emph{categorical}. In ordinal multilayer networks, interlayer edges exist only between layers that are adjacent to each other with respect to some criterion (e.g., temporal ordering). By contrast, categorical multilayer networks include interlayer edges between replica nodes from every pair of layers. For the sake of simplicity, we illustrate \muxviz\, using interlayer edges of weight 1 in the present paper. In general, how to choose such weights is an open research question. See the discussions in Ref.~\cite{Bassett2013Robust,kivela2013multilayer}.
}

\begin{figure}[!b]
\centering\includegraphics[width=16cm]{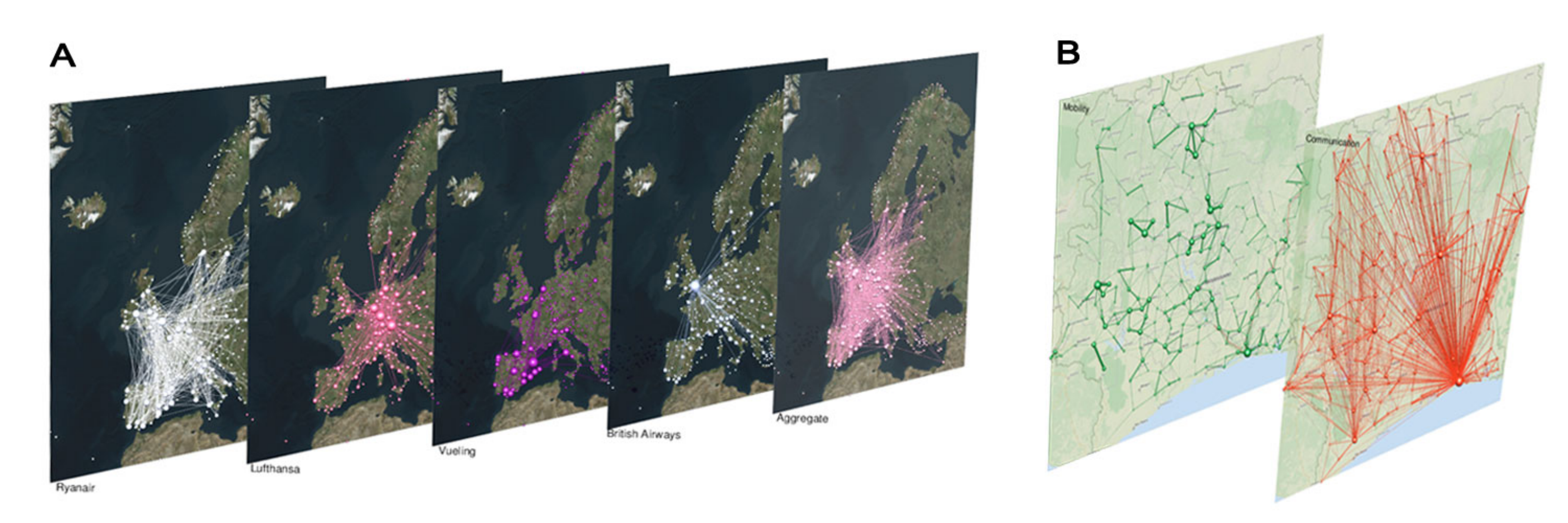} 
\caption{\textbf{Multilayer networks embedded in geographical regions}. \textbf{(A)} Network of European airports, where each layer represents a different airline \cite{cardillo2013emergence}. \textbf{(B)} Network of mobility and communication in the Ivory Coast, where nodes are geographical districts \cite{lima2013exploiting}. We used \muxviz\, to visualize these data sets.
}
\label{fig:fig6suppl}
\end{figure}

\rev{
For instance, let us examine the genetic-interaction and profile-correlation networks of a cell as different layers for a multilayer network. Such networks were aggregated into a single network in Ref.~\cite{costanzo2010genetic}. In Fig.\,\ref{fig:fig1suppl}{\bf A}, we show multilayer visualizations that we created using \muxviz.  Other representations are also possible \cite{kivela2013multilayer}.  For example, when representing this data as an edge-colored multigraph, we ``color'' edges according to the type of relationship that they represent (see Fig.\,\ref{fig:fig1suppl}{\bf B}). In Fig.\,\ref{fig:fig1suppl}{\bf C} and Fig.\,\ref{fig:fig1suppl}{\bf D}, we show the two visualizations for the connectome of \emph{Caenorhabditis elegans}. In this example, each layer corresponds to a different type of synaptic connection \cite{chen2006wiring}. 
}

\rev{
In panels {\bf (A)} and {\bf (C)} of Fig.\,\ref{fig:fig1suppl}, we use a layout in which the positions of the nodes are the same in each layer. We determine the positions of nodes by combining two of the standard force-directed algorithms available in \muxviz\, and applying them to an aggregated network that we obtained by summing the corresponding entries of the adjacency matrices of the individual layers. Specifically, we first apply the Fruchterman-Reingold algorithm \cite{fruchterman1991graph} to the aggregated network and then use the output of this algorithm as a seed layout for the Kamada-Kawai algorithm \cite{kamada1989algorithm} to achieve a better spatial separation of nodes in the final layout. The \muxviz\, software also allows other layout choices. For example, the layout of each layer can be independent, or one can determine node locations using any individual layer or an aggregation over any subset of layers.
}

\rev{One can also use \muxviz\, for a large variety of other analyses and visualizations. For example, as we illustrate in Fig.\,\ref{fig:fig6suppl},  \muxviz\, can account for spatial information by creating visualizations of multilayer networks that are embedded in geographical regions.
}


\rev{\subsection{Compression of layers and reducibility dendrogram}}\label{supp:note:2}

\rev{
An important open question is the determination of how much information is necessary to accurately represent the structure of multilayer systems and whether it is possible to aggregate some layers without loss of information. It was shown recently that it is possible to \rev{compress} the number of layers in multilayer networks in a way that minimizes information loss by using an information-theoretic approach \cite{dedomenico2014reducibility}. The methodology of ~\cite{dedomenico2014reducibility}, which we implemented in \muxviz, amounts to a tradeoff (which is ``optimal'' in some sense) between accuracy and complexity.  Alternatively, users of \muxviz\, can implement alternative methods based on different notions of ``minimal information loss''.
}

\rev{
The \rev{compression} procedure from ~\cite{dedomenico2014reducibility} proceeds as follows. For each pair of layers in the original multilayer network, \muxviz\, calculates the quantum Jensen--Shannon (JS) divergence \cite{majtey2005jensen}. This estimates the similarity between two networks based on their Von Neumann entropy \cite{braunstein2006laplacian}. By definition, the quantum Jensen--Shannon divergence is symmetric and its square root, which is usually called the Jensen--Shannon distance, satisfies the properties of a metric \cite{briet2009properties}. One can use the JS distance to quantify the distance in terms of information gain (or loss) between the normalized Laplacian matrices that are associated to two distinct networks \cite{dedomenico2014reducibility}.
}

\rev{
One places the distances between every pair of layers as the components of a matrix, and one can then perform hierarchical clustering \cite{dataclustering} using any desired method to produce a dendrogram that indicates the relatedness of the information in the different layers. In \muxviz, we have included several methods for hierarchical clustering (e.g., Ward, McQuitty, single, complete, average, median, and centroid linkage clusterings).  We show an example of such a ``reducibility dendrogram'' in panel {\bf (D)} of Figs.\,\ref{fig:fig2-1suppl}, \ref{fig:fig3-1suppl}, \ref{fig:fig4-1suppl}, and \ref{fig:fig5-1suppl}.  A reducibility dendrogram results from a step-by-step merging of a set of layers in a multilayer network, and we calculate a quality function based on the relative Von Neumann entropy to estimate information gain (or loss) at each step \cite{dedomenico2014reducibility}.  To obtain a reduced version of the original multilayer network, we stop the merging procedure at the level of the hierarchy that maximizes the relative entropy. 
}

\subsection{Annular visualization of multilayer information}\label{supp:note:3}

\rev{
It is a challenging problem to represent, visualize, and analyze the wealth of information encoded in the multilayer structure of networks in a compact way.  Preserving more information by using multilayer networks rather than ordinary networks complicates the visualization and analysis even further.  However, this complication is necessary, because otherwise one might end up with misleading or even incorrect results \cite{kivela2013multilayer}.  We developed the \muxviz\, software to help address these challenges. To summarize all of the information obtained from multilayer-network calculations in a compact way, \muxviz\, includes an annular visualization that facilitates the ability to capture patterns and deduce qualitative information about multilayer data.
}

\rev{
To give a concrete example, many researchers are interested in ranking the relative importance of nodes (and other network structures), which traditionally is accomplished using various ``centrality'' measures.  Centralities have been calculated for single-layer networks for several decades \cite{Wasserman1994Social,newman2010}, and numerous notions of centrality are now also available for multilayer networks \cite{dedomenico2013centrality,kivela2013multilayer}.  It is therefore necessary to develop visualization tools that make it possible to compare such a wealth of diagnostics to each other in a compact, meaningful way.  For example, it is often worthwhile to focus attention on one descriptor and compare the values obtained in each layer separately to the values obtained from the multilayer network and its aggregations. This is easy to do using the \muxviz\, software.
}

\rev{
We will now illustrate our annular visualization (see Figs.\,\ref{fig:fig2-2suppl}, \ref{fig:fig3-2suppl}, \ref{fig:fig4-2suppl}, and \ref{fig:fig5-2suppl}) using the example of multilayer centrality measures.  Suppose that we have different arrays of information, where one should think of each array as having resulted from the calculation of some centrality diagnostic on a multilayer network. We visualize each array using a ring. The angle indicates node identity (regardless of the layer or layers in which it occurs).  We bin the centrality values---e.g., either linearly or logarithmically---and we assign a color to each bin to encode its value. Both the type of binning and the color scheme are customizable in \muxviz.  We place the rings concentrically, and one can determine both the ring order and ring thicknesses according to any desired criteria.  For example, in the visualizations in the present paper, we determine the thickness of each ring according to its information content (which we quantify using the Shannon information entropy of the distribution of the values): thinner rings have less information. Users can customize the order of the rings; in \muxviz's default setting, it is determined automatically via hierarchically clustering. The \muxviz\, software calculates a measure of correlation (e.g., Pearson, Spearman, or JS divergence) between each pair of descriptors to obtain a set of pairwise distances, which we then hierarchically cluster to group the rings. This clustering procedure determines the order of the rings to try to maximize the readability of the annular plot.
}

\rev{One can also use the same principles when fixing some centrality descriptor and letting the rings correspond to the layers in a network, the multilayer network, and an aggregated network (see Section \ref{anal}). Such a plot might help to reveal, for instance, if the centrality of nodes in a multilayer network is primarily due to their centrality in a specific layer or if the aggregated network is a good proxy for the multilayer structure.
}


\section{Analyses of empirical multilayer networks} \label{anal}

\begin{figure}[!t]
\centering\includegraphics[width=16cm]{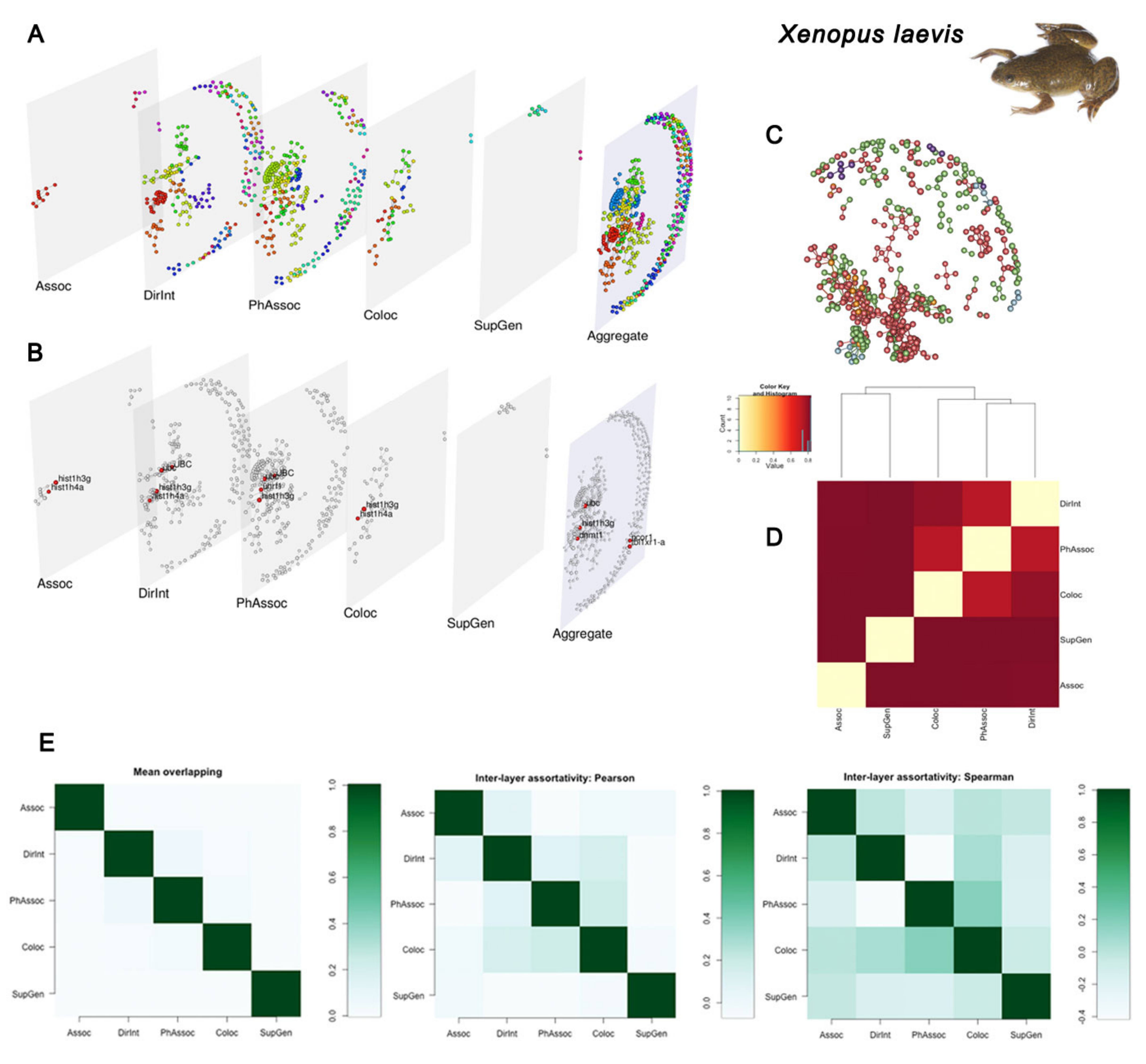} 
\caption{\textbf{Multilayer analysis of a \emph{Xenopus laevis} genetic-interaction network.}  See Section\,\ref{supp:note:4} for details about each panel.
[In this figure and all subsequent figures, we have purposely kept font sizes at \muxviz's default level rather than increasing them.]
}
\label{fig:fig2-1suppl}
\end{figure}

To demonstrate the ability of \muxviz\, to analyze and visualize multilayer networks, we consider different types of genetic interactions for organisms in the Biological General Repository for Interaction Datasets\cite{stark2006biogrid} (BioGRID, \url{thebiogrid.org}), a public database that archives and disseminates genetic and protein interaction data from humans and model organisms. BioGRID currently includes more than 720,000 interactions that have been curated from both high-throughput data sets and individual focused studies using over 41,000 publications in the primary literature. We use BioGRID 3.2.108 (updated 1 Jan 2014). In this section, we focus on \emph{Xenopus laevis} and show a network visualization in Fig.\,\ref{fig:fig2-1suppl}\textbf{C}. We give results of computations using \muxviz\, in the other panels of Fig.\,\ref{fig:fig2-1suppl}. See Section\,\ref{supp:note:4} for other examples. 

\begin{figure}[!t]
\centering\includegraphics[width=16cm]{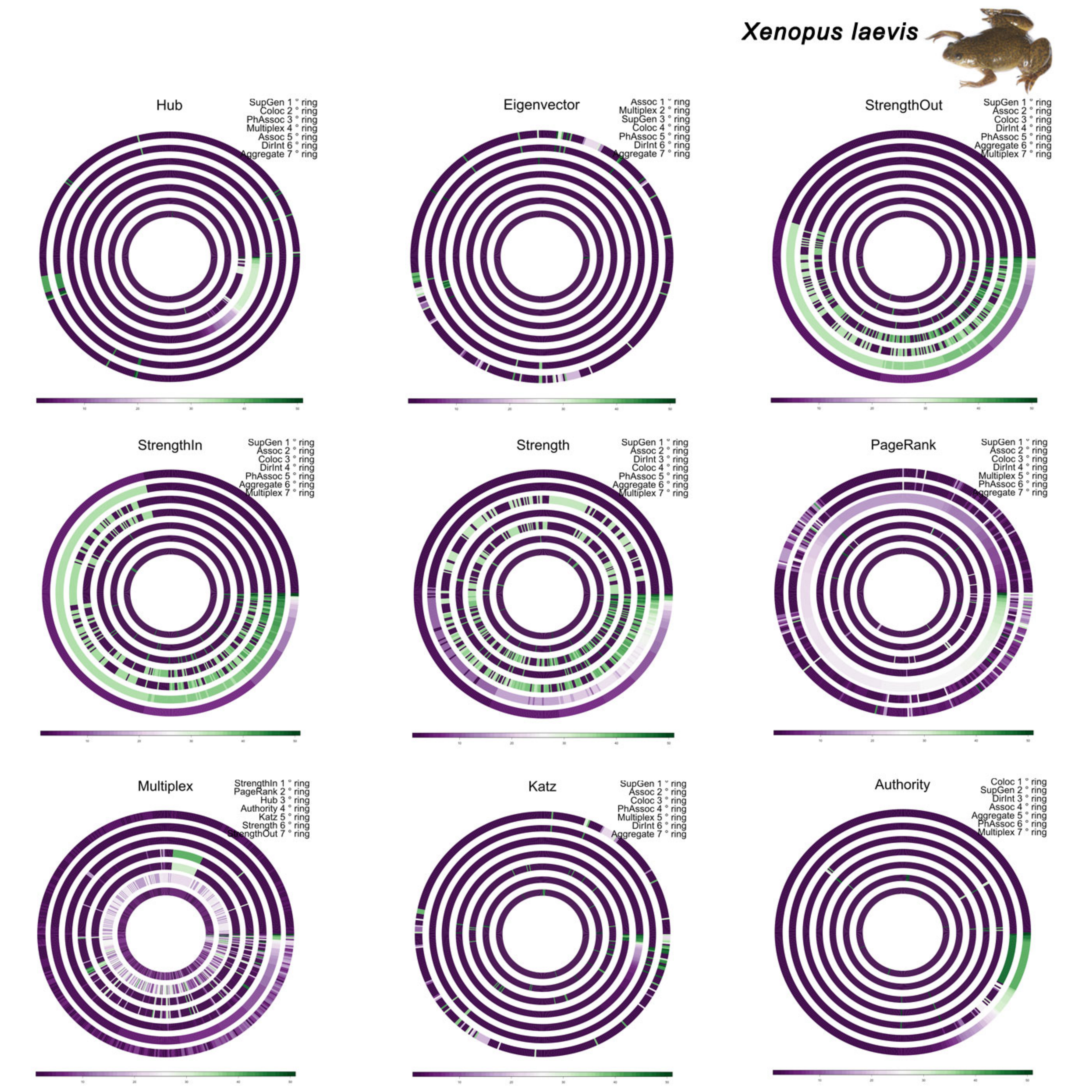} 
\caption{\textbf{Multilayer analysis of a \emph{Xenopus laevis} genetic-interaction network.}  See Section\,\ref{supp:note:4} for details about each panel.
}
\label{fig:fig2-2suppl}
\end{figure}

One can examine the global organization of nodes into modules (i.e., ``communities'') through an algorithmic calculation of community structure \cite{Porter2009,fortunato2010community}. For example, one can obtain dense communities in multilayer networks by optimizing a multilayer generalization of the modularity quality function \cite{mucha2010community}. To do this, one takes into account both intralayer and interlayer edges, and one seeks densely connected sets of nodes (i.e., communities) that are sparsely connected to each other as compared to some multilayer random-graph (null) model \cite{mucha2010community,Bassett2013Robust,kivela2013multilayer}. See Fig.\,\ref{fig:fig2-1suppl}\textbf{A} for a visualization of communities in \emph{Xenopus laevis} and Section\,\ref{supp:note:4} for other examples. 

As we discussed previously, one can quantify the importance of a node by using various diagnostics to measure ``centrality''. One calculates such a centrality (and a corresponding rank order) for each node by using multilayer generalizations of centrality measures \cite{dedomenico2013mathematical,dedomenico2013centrality,kivela2013multilayer}. The software \muxviz\, has tools for calculating multilayer generalizations of several different types of centrality (e.g., degree, eigenvector \cite{bonacich1972factoring}, hub and authority \cite{kleinberg1999authoritative}, PageRank \cite{brin1998anatomy}, and Katz \cite{katz1953new}) either for an entire multilayer network or for each layer separately.  As we illustrate in Fig.\,\ref{fig:fig2-1suppl}\textbf{B}, centrality values (as well as other network measures) can be very different in multilayer networks than in their corresponding aggregations. Such results influence how one should interpret calculations of network measures for, e.g., which genes or proteins are most important for activating or suppressing a given biological processes \rev{or which people are most important in social networks.} The data in question is multilayer, so the analysis of such data must take multilayer features into account.

Researchers are often also interested in considering a ``\rev{compressed} version'' of multilayer data sets that preserve as much information as possible without altering the primary descriptors. For such scenarios, it is possible to use the \rev{compression procedure} discussed in Section\,\ref{supp:note:2} to identify the layers of a multilayer network that are providing redundant information \cite{dedomenico2014reducibility} \rev{(see Fig.\,\ref{fig:fig2-1suppl}\textbf{D})}. 

In Fig.\,\ref{fig:fig2-1suppl}\textbf{E}, we show three correlation measures for multilayer networks: (left) mean edge overlap, (center) degree-degree Pearson correlation coefficient, and (right) degree-degree Spearman correlation coefficient. In this example, the degree-degree Pearson and Spearman correlation coefficients between layers quantify the tendency of nodes to be hubs in different layers simultaneously. The \muxviz\, software can include additional correlation measures, and it is easy for users to implement other diagnostics \cite{nicosia2013correlations}.

To summarize all of the information that one obtains from calculations like the ones above in a compact figure, we use an annular visualization (see Section\,\ref{supp:note:3}) that facilitates the ability to capture patterns to deduce qualitative information about multilayer data. In Fig.\,\ref{fig:fig2-2suppl} (see the panel labelled ``Multiplex''), we show an example for centrality diagnostics, which measure the importance of nodes in various ways. Each ring indicates a centrality measure, and the angle determines the identity of a node in a network, regardless of the layer(s) in which it exists. One can use the same principles when fixing some centrality descriptor and letting the rings correspond to the layers in a network, the multilayer network, and an aggregated network (see the other panels in Fig.\,\ref{fig:fig2-2suppl}). For the case of layers, one calculates a centrality measure for each layer separately without accounting for multilayer structure. For instance, it is evident that rings 3 (``DirInt'' layer) and 5 (``PhAssoc'' layer) are negatively correlated in the case of strength centrality because nodes tend to have opposite colors, whereas rings 6 (aggregated network) and 7 (multiplex network) are positively correlated, as expected for strength centrality. Our annular representation makes it easy to see similarity (or dissimilarity) in rank orderings according to different diagnostics.  For example, their patterns reveal that physical association and direct interaction are dominant and determine the multilayer strength \rev{in the depicted example.} In other cases (see Section\,\ref{supp:note:4}), the ranking by some centrality measure in the multilayer network is poorly correlated to the ranking in either an aggregated network or in individual layers separately. \rev{This underscores the value of using a multilayer framework for the calculation of the most central proteins (and, more generally, for determining which entities in many complex systems are most important).}

\begin{figure}[!t]
\centering\includegraphics[width=16cm]{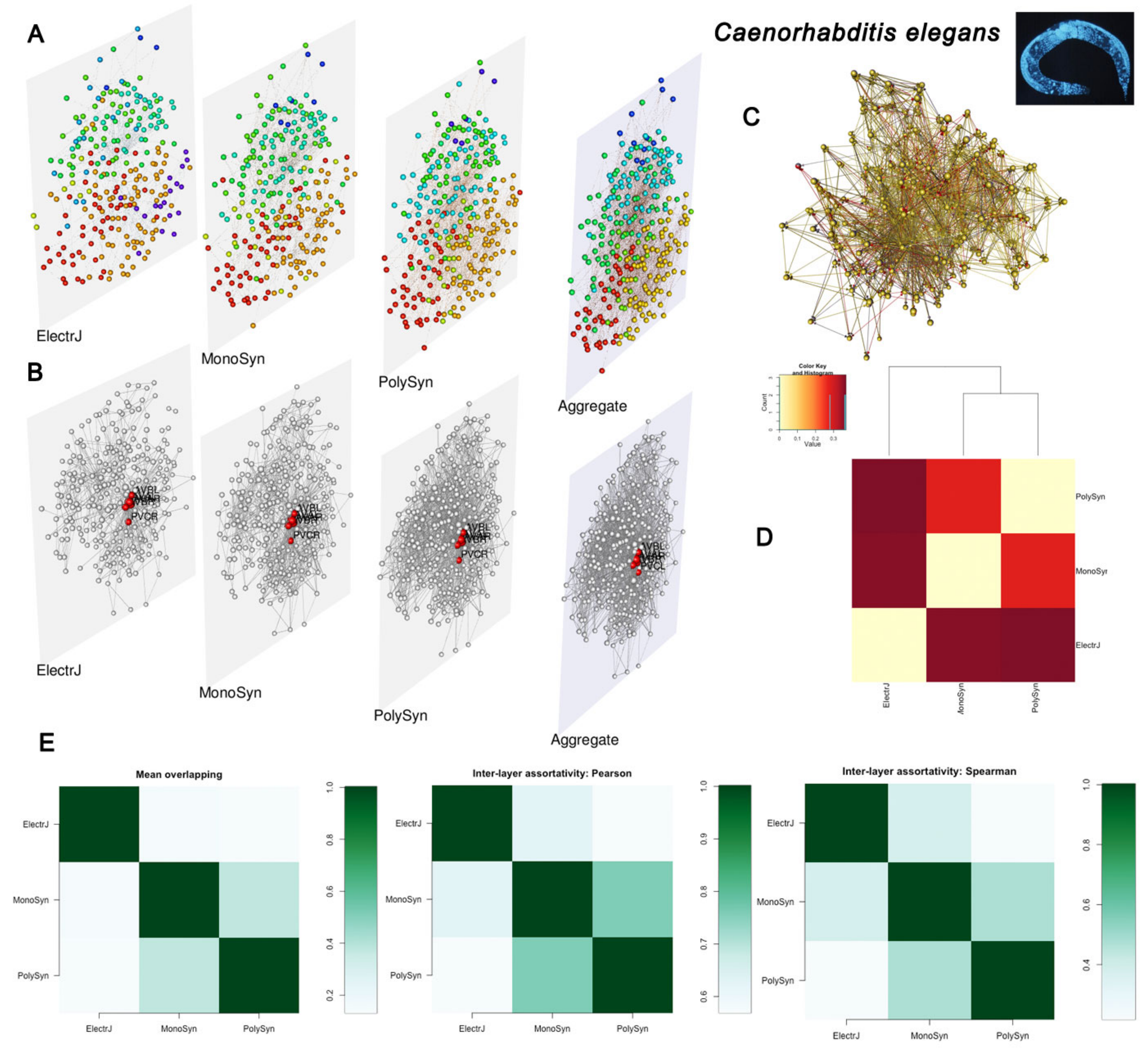} 
\caption{\textbf{Multilayer analysis of a \emph{Caenorhabditis elegans} connectome.}  See Section\,\ref{supp:note:4} for details about each panel.}
\label{fig:fig3-1suppl}
\end{figure}
\begin{figure}[!t]
\centering\includegraphics[width=16cm]{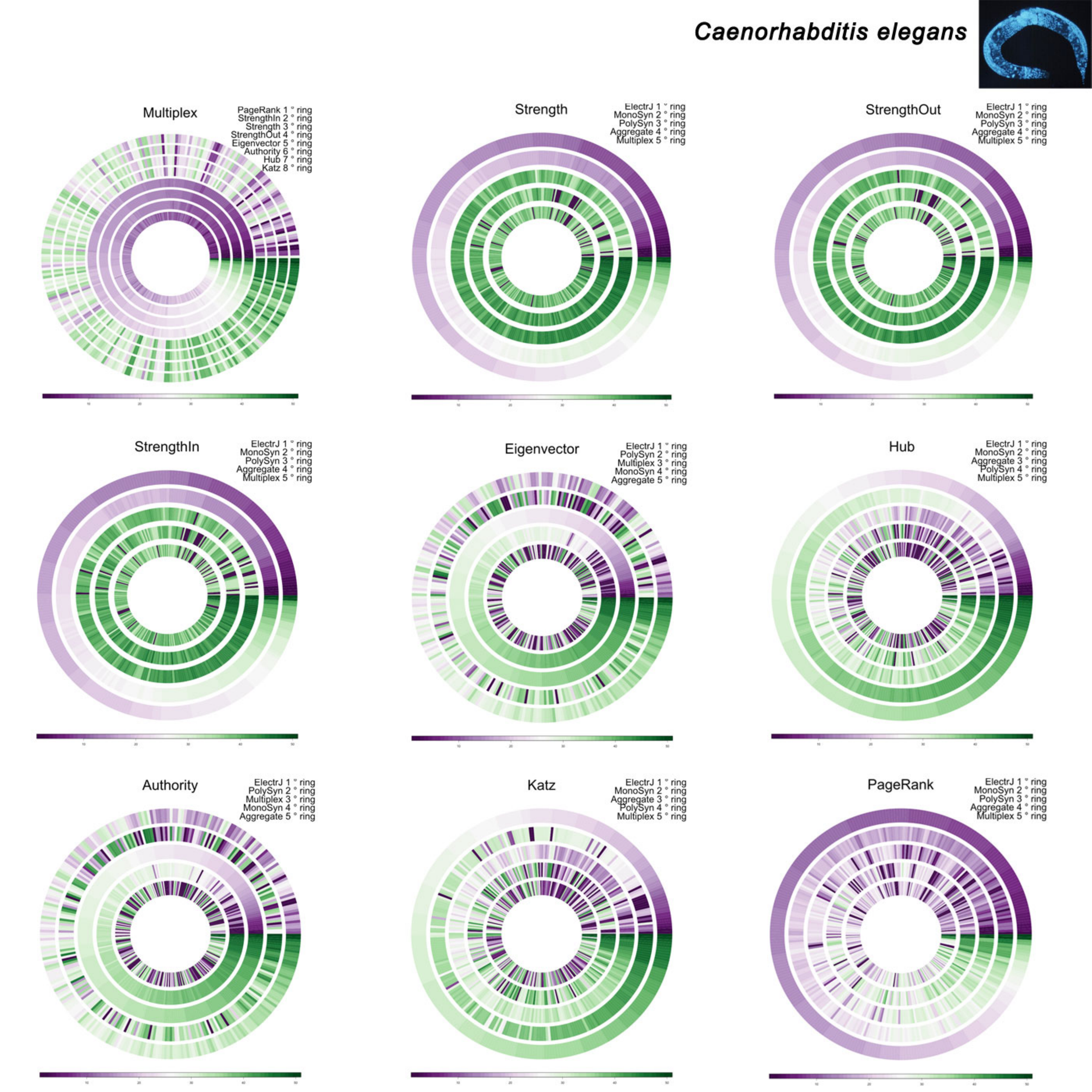} 
\caption{\textbf{Multilayer analysis of a \emph{Caenorhabditis elegans} connectome.}  See Section\,\ref{supp:note:4} for details about each panel.}
\label{fig:fig3-2suppl}
\end{figure}
\begin{figure}[!t]
\centering\includegraphics[width=16cm]{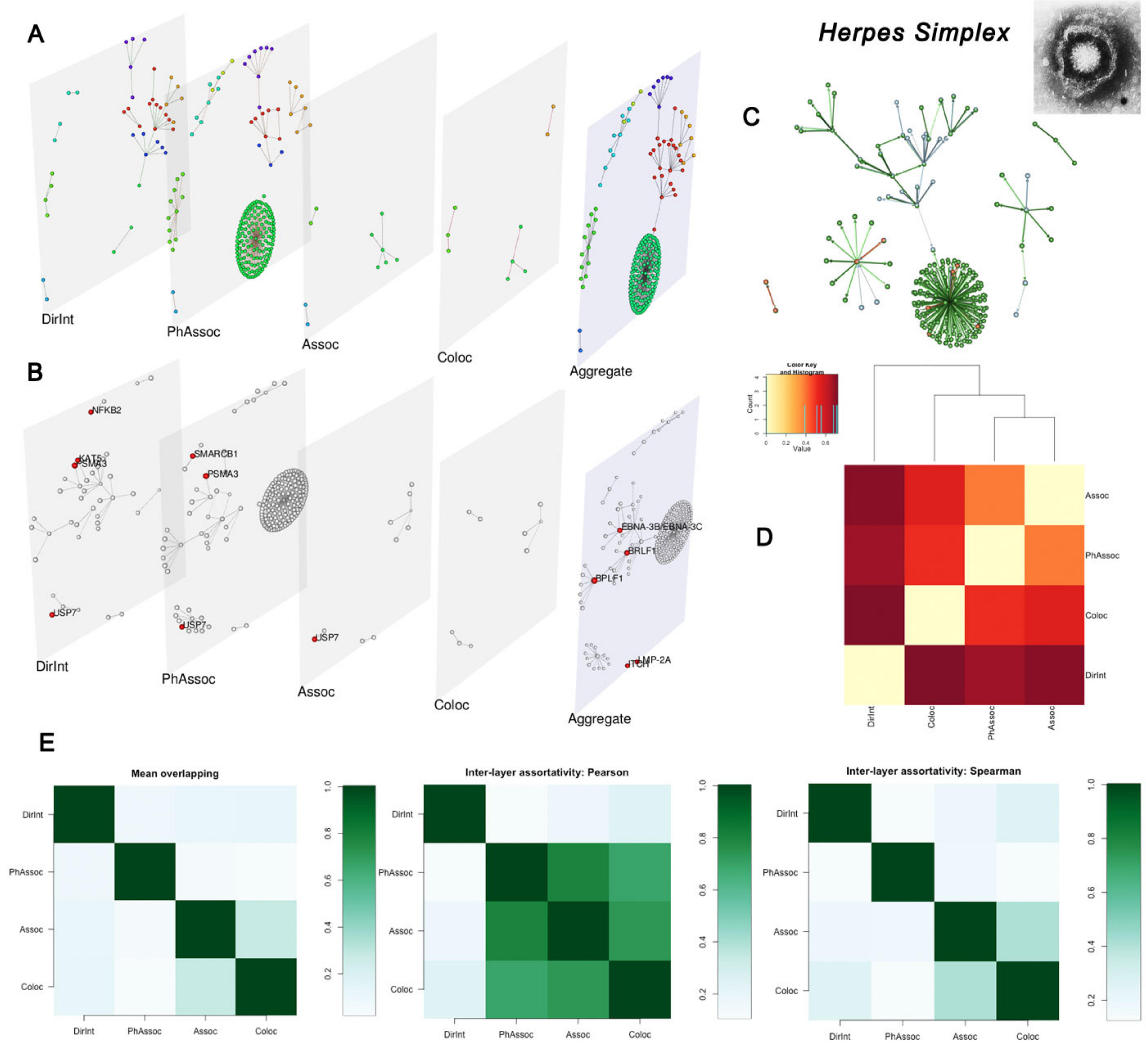} 
\caption{\textbf{Multilayer analysis of a \emph{Herpes simplex} genetic-interaction network.}  See Section\,\ref{supp:note:4} for details about each panel.}
\label{fig:fig4-1suppl}
\end{figure}
\begin{figure}[!t]
\centering\includegraphics[width=16cm]{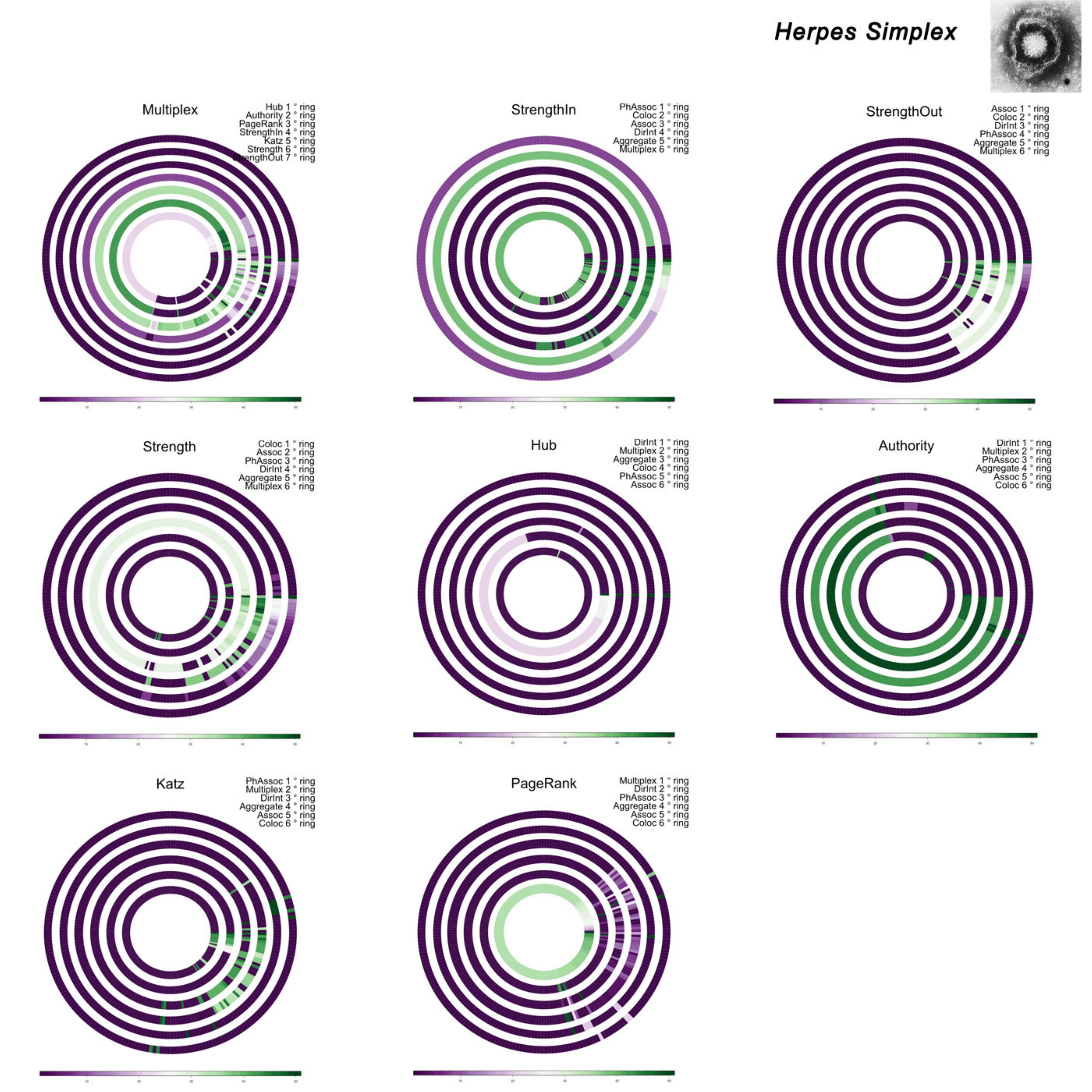} 
\caption{\textbf{Multilayer analysis of a \emph{Herpes simplex} genetic-interaction network.}  See Section\,\ref{supp:note:4} for details about each panel. Note that we do not show eigenvector centrality because one layer consists of a directed acyclic graph (for which eigenvector centrality is unilluminating \cite{newman2010}).
}
\label{fig:fig4-2suppl}
\end{figure}
\begin{figure}[!t]
\centering\includegraphics[width=16cm]{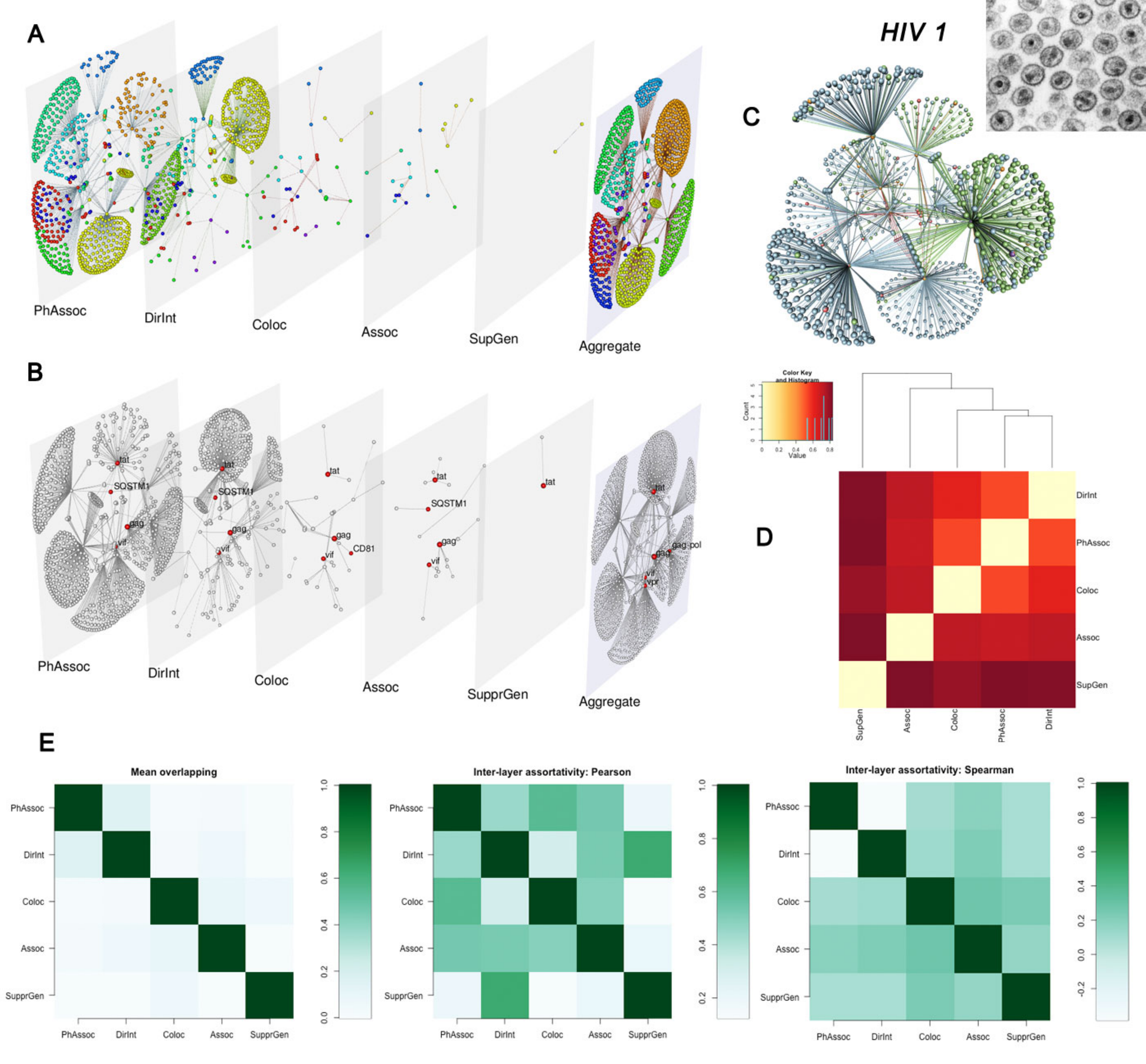} 
\caption{\textbf{Multilayer analysis of \emph{HIV-1} genetic-interaction network.} See Section\,\ref{supp:note:4} for details about each panel.}
\label{fig:fig5-1suppl}
\end{figure}
\begin{figure}[!t]
\centering\includegraphics[width=16cm]{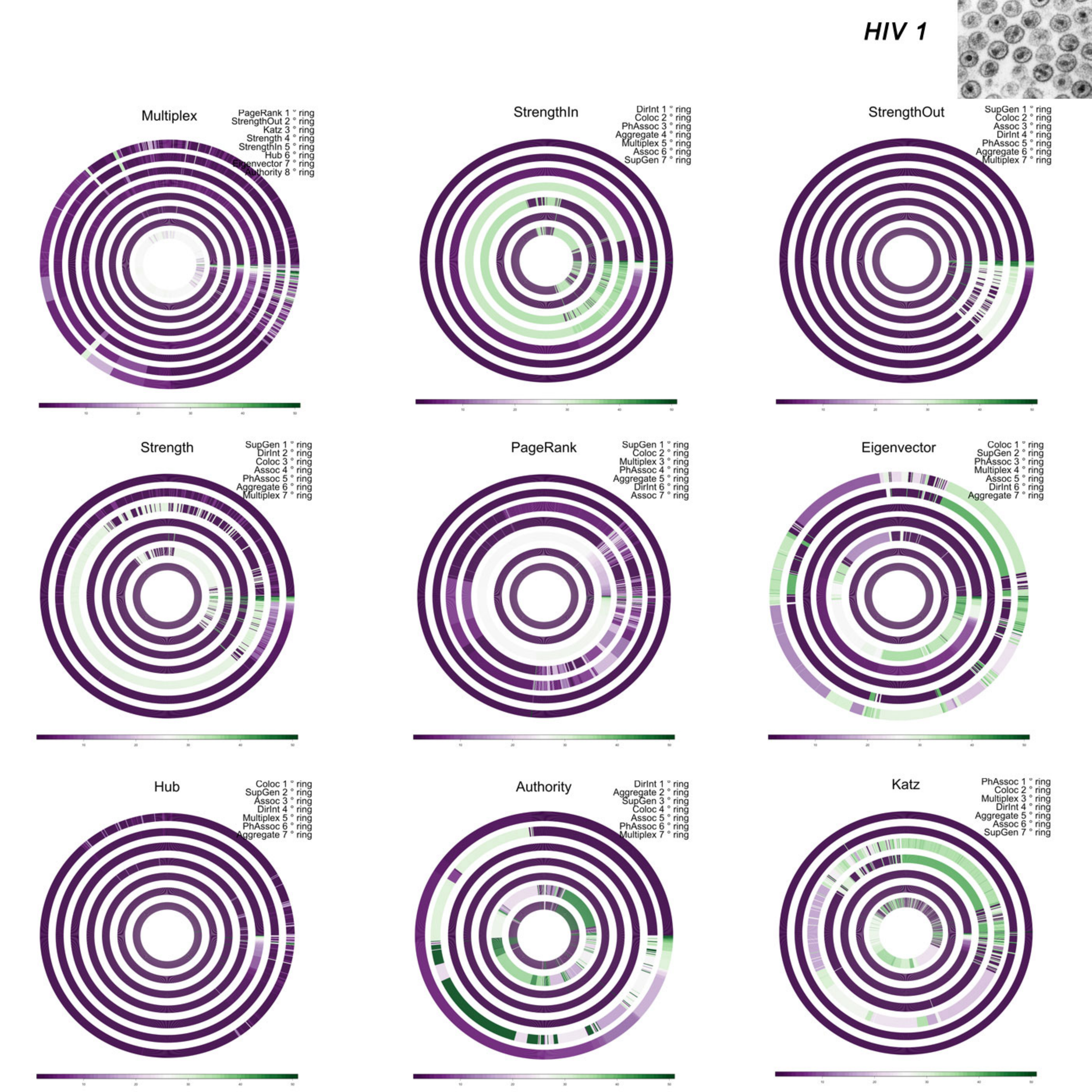} 
\caption{\textbf{Multilayer analysis of \emph{HIV-1} genetic-interaction network.} See Section\,\ref{supp:note:4} for details about each panel.}
\label{fig:fig5-2suppl}
\end{figure}

\rev{
\subsection{Analysis of other empirical multilayer networks}\label{supp:note:4}

In this section, we present multilayer analyses of three additional biological systems to illustrate the power of \muxviz. We examine the following examples:
\begin{itemize}
\item \emph{Caenorhabditis elegans} connectome (see Figs.\,\ref{fig:fig3-1suppl} and \ref{fig:fig3-2suppl});
\item \emph{Herpes simplex} genetic-interaction network (see Figs.\,\ref{fig:fig4-1suppl} and \ref{fig:fig4-2suppl});
\item \emph{HIV-1} genetic-interaction network (see Figs.\,\ref{fig:fig5-1suppl} and \ref{fig:fig5-2suppl}).
\end{itemize}

As for the case of \emph{Xenopus laevis}, we include two figures for each example. In the first set of figures (see \ref{fig:fig3-1suppl}, \ref{fig:fig4-1suppl}, and \ref{fig:fig5-1suppl}), we show the following information:
\begin{itemize}
\item \textbf{Panel A}: Multilayer community structure from modularity maximization \cite{mucha2010community}. The color of each node encodes its community assignment in a multilayer-network visualization.  For comparison, we also show the results (and corresponding visualization) of community detection on an aggregated network, which we obtain by summing the corresponding intralayer edge weights of all layers. (In other words, if $A_{ijs}$ gives the edge weight between nodes $i$ and $j$ on layer $s$, then we obtain an aggregated edge weight $W_{ij}$ between nodes $i$ and $j$ by summing over $s$.)
\item \textbf{Panel B:} Multilayer PageRank centrality \cite{dedomenico2013centrality}. We again use a multilayer-network visualization.  We label the top five nodes from a ranking according to multilayer PageRank centrality.  For comparison, we also show the results of PageRank centrality calculations on the aforementioned aggregated network.
\item \textbf{Panel C:} Edge-colored multigraph visualization of the network.  We color edges according to the layer to which they belong.  We color the nodes according to their layer (or layers); if a node exists on multiple layers, then we distribute its corresponding colors evenly.
\item \textbf{Panel D:} \rev{Compressibility} analysis and corresponding reducibility dendrogram \cite{dedomenico2014reducibility}. We show the distance matrix and the corresponding dendrogram, which we obtain using Ward hierarchical clustering. 
\item \textbf{Panel E:} Measures of correlation between layers: (left) mean edge overlap, (center) degree-degree Pearson correlation coefficient, and (right) degree-degree Spearman correlation coefficient. 
\end{itemize}

In the second set of figures (see Figs.\,\ref{fig:fig3-2suppl}, \ref{fig:fig4-2suppl}, and \ref{fig:fig5-2suppl}), we show the annular visualization for the centrality descriptors:
\begin{itemize}
\item In panels titled ``Multiplex'', we consider the multilayer network. Each ring corresponds to a different centrality descriptor.
\item In the other panels, we consider a specific centrality descriptor (which we specify in the title of the panel). Each ring encodes the values of that descriptor, which we calculate in each layer separately.  We also include rings for the calculation of the corresponding centrality diagnostic in the multilayer network and in its aggregation to a single-layer weighted network.
\end{itemize}
We specify the order of the rings in the list of labels on the right of each plot.  In each case, the top label refers to the innermost ring and the bottom label refers to the outermost ring.
}

\section{Conclusion}

In the current era of ``big data'', there is now an intense deluge of multilayer data.  To avoid throwing away important information or obtaining misleading results, it is increasingly crucial to use methods that exploit multilayer structure.  In this paper, we present new software and associated methodology that exploits the new paradigm of multilayer networks, and we illustrate how it can be used to analyze and visualize several examples.
Our software, \muxviz, provides an open-source framework for the analysis of multilayer networks. Additionally, the modular structure of \muxviz\, --- along with its open-source license --- makes it easy to add new methods. \rev{Moreover, although we have focused on examples of biological networks, \muxviz\, is clearly also useful for multilayer networks from any other setting.} As we illustrate in Fig.\,\ref{fig:fig6suppl}, it can even be overlaid over spatial information.

\section*{Acknowledgements}

All authors were supported by the European Commission FET-Proactive project PLEXMATH (Grant No. 317614). AA also acknowledges financial support from the Generalitat de Catalunya 2009-SGR-838, the ICREA Academia, and the James S.\ McDonnell Foundation. MAP acknowledges a grant (EP/J001759/1) from the EPSRC. We thank Serafina Agnello for support with graphics.

\clearpage
\bibliography{muxviz_final}
%


\appendix

\setcounter{figure}{0}


\section{Technical Details About \muxviz}\label{supp:note:5}

We developed \muxviz\, using R (\url{http://www.r-project.org/}), a free and widely-adopted framework for statistical computing, and GNU Octave (\url{https://www.gnu.org/software/octave/}), an open-source high-level interpreted language that is intended primarily for numerical computations. The Octave language is very similar to the proprietary environment {\sc Matlab} (\url{http://www.mathworks.es/products/matlab/}), and one can import the code to {\sc Matlab} in a straightforward manner. The \muxviz\, software requires R 3.0.2 (or above) and Octave 3.4.0 (or above).

The \muxviz\, framework is a free and open-source package for the analysis and the visualization of multilayer networks. It is released under GNU General Public License v3 (\url{https://www.gnu.org/copyleft/gpl.html}) and exploits R to provide an easy and accessible user interface for the visualization of networks, the calculation of network diagnostics, and the visual representation of the results of calculations. Specifically, R allows the construction of a graphical user interface (GUI), which can be used either locally (client-side software) or via the internet (remote Web server), and an Octave library that we developed performs calculations of matrices and tensors.

Using \muxviz\, is simple and does not require any programming skill; one can do all computations and visualization via the user interface. Additionally, because of \muxviz's modular structure, users can also create their own modules for calculating new diagnostics and for customizing visual representations.

The \muxviz\, framework allows both two-dimensional and three-dimensional visualization of networks. The latter exploits OpenGL technology, so users can interactively change the perspective and navigate the network. We show representative static snapshots of such interactive visualizations in Figs.\,\ref{fig:fig1suppl}\textbf{B} and \textbf{D} and in panel \textbf{C} of Figs.\,\ref{fig:fig2-1suppl}, \ref{fig:fig3-1suppl}, \ref{fig:fig4-1suppl}, and \ref{fig:fig5-1suppl}.

\end{document}